\documentclass[prd,twocolumn,tightenlines]{revtex4-1}
\usepackage{tipa}
\usepackage{mathrsfs}
\usepackage{amsmath}
\usepackage{amssymb}
\usepackage{graphicx}

\newcommand{\be}{\begin{eqnarray}}
\newcommand{\ee}{\end{eqnarray}}

 \newcommand{\gsim}{\mathrel{\hbox{\rlap{\lower.55ex \hbox {$\sim$}}
                   \kern-.3em \raise.4ex \hbox{$>$}}}}
\newcommand{\lsim}{\mathrel{\hbox{\rlap{\lower.55ex \hbox {$\sim$}}
                   \kern-.3em \raise.4ex \hbox{$<$}}}}

\begin{document}

\title{ Centrality dependence of charmed-meson photoproduction in ultrarelativistic heavy ion collisions }

\author{Gong-Ming Yu$^{1}$}
\email{ygmanan@163.com}

\author{Yan-Bing Cai$^{2}$}

\author{Zhen Bai$^{1}$}

\author{Qiang Hu$^{1}$}

\author{Jian-Song Wang$^{1,}$}
\email{jswang@impcas.ac.cn}

\address{ $^{1}$Institute of Modern Physics, Chinese Academy of Sciences, Lanzhou 730000, China\\$^{2}$Department of Physics, Yunnan University, Kunming 650091, China }





\begin{abstract}
We calculate the centrality dependence of inclusive cross section of large-$p_{T}$ charmed-meson ($D^{0}$, $D^{*}$, $D^{*+}$, and $D_{s}^{+}$) from heavy quark fragmentation by the hard photoproduction processes for the nucleus-nucleus collisions in ultrarelativistic heavy ion collisions. The numerical results indicate that the modification of the hard photoproduction processes cannot be negligible for the charmed-meson production in Au-Au collisions at Relativistic Heavy Ion Collider (RHIC) and Pb-Pb collisions at Large Hadron Collider (LHC).


\end{abstract}

\maketitle


\section{Introduction}

Recent measurements of charmed-meson ($D^{0}$, $D^{*}$, $D^{*+}$, and $D_{s}^{+}$) production in high energy hadronic collisional experiment reported by PHENIX Collaboration\cite{1} and STAR collaboration\cite{2,3,4,5,6,7,8} at Relativistic Heavy Ion Collider (RHIC) as well as ALICE Collaboration\cite{9,10,11,12,13,14,15,16,17,18,19,20}, ATLAS Collaboration\cite{21,22}, and LHCb Collaboration\cite{23,24} at Large Hadron Collider (LHC) have shown that inclusive heavy-quark production provides us with an important tools for the test of the predictions of perturbative quantum chromodynamics (pQCD)\cite{25,26}, and the evolution of charm quark in hot partonic medium can be considered as the probe of the quark-gluon plasma produced in ultrarelativistic heavy ion collisions\cite{27,28,29,30,31,32,33,34,35,36,37}. Indeed, many phenomenological models such as the color glass condensate (CGC) effective theory\cite{38}, general-mass variable-avor-number scheme\cite{39,40,41}, $k_{T}$-factorization approach\cite{42,43}, fragmentation approach\cite{44,45,46}, Parton-Hadron-String Dynamics (PHSD) transport approach\cite{47,48}, heavy quark recombination mechanism\cite{49,50}, two-component HYDJET++ model\cite{51,52}, gluon splitting with Langevin transport model\cite{53}, POWLANG transport model\cite{54}, double-parton scattering (DPS) mechanisms\cite{55}, quark coalescence model\cite{56}, and single- and central-diffractive mechanisms in the Ingelman-Schlein model\cite{57} have been proposed for the calculation of charmed-meson production.

Although considerable efforts both in experiment and theory, the charmed-meson production mechanism is still not fully understood. In the present work, we extend the hard photoproduction mechanism\cite{58,59,60,61,62,63} that plays a fundamental role in the electron-proton deep inelastic scattering at the Hadron Electron Ring Accelerator (HERA) to the charmed-meson production in Au-Au collisions at RHIC and Pb-Pb collisions at LHC. At high energies, the charged partons of incident nucleus can emit high energy photons (hadron-like photons) in ultrarelativistic nucleus-nucleus collisions. The hard photoproduction processes may be direct and resolved processes that are sensitive to the gluon distribution in the incident nucleus. In the hard direct photoproduction processes, the high energy photon emitted from the charged parton of the incident nucleus interacts with the parton of another incident nucleus. In the hard resolved photoproduction processes, the uncertainty principle allows the high energy hadron-like photon emitted from the charged parton of the incident nucleus for a short time to fluctuate into a quark-antiquark pair which then interacts with the partons of another incident nucleus.

The paper is organized as follows. In Sec. II we present the photoproduction of charmed-meson at RHIC and LHC energies. The numerical results for large-$p_{T}$ charmed-meson in Au-Au collisions at RHIC and Pb-Pb collisions at LHC are also plotted. The conclusion is given in Sec. III.


\section{General formalism}

The centrality dependence of differential cross section for inclusive charmed-meson hadroproduction from the initial parton hard scattering processes (LO) in hadronic collisions can be expressed as
\begin{eqnarray}\label{y1}
\frac{dN^{LO}}{dp_{T}^{2}dy}&=&\int d^{2}bd^{2}rdx_{a}dx_{b}T_{A}(\mathbf{r})T_{B}(|\mathbf{r}-\mathbf{b}|)\nonumber\\[1mm]
&&\times f_{a/A}(x_{a},Q^{2},\mathbf{r})f_{b/B}(x_{b},Q^{2},|\mathbf{r}-\mathbf{b}|)\nonumber\\[1mm]
&&\times \frac{x_{a}x_{b}}{x_{a}x_{b}-\tau}\frac{d\hat{\sigma}}{d\hat{t}}(ab\rightarrow cd)\frac{D_{H/c}(z_{c})}{z_{c}},
\end{eqnarray}
where the variables $x_{a}$ and $x_{b}=(x_{a}x_{2}-\tau)/(x_{a}-x_{1})$ are the momentum fractions of the partons, $z_{c}$ is the momentum fraction of the final charmed-meson, $x_{1}=\frac{1}{2}(x_{T}^{2}+4\tau)^{1/2}\exp(y)$, $x_{2}=\frac{1}{2}(x_{T}^{2}+4\tau)^{1/2}\exp(-y)$, $x_{T}=2p_{T}/\sqrt{s}$, $\tau=(M/\sqrt{s})^{2}$, and $M$ is the mass of the charmed-meson; $\frac{d\hat{\sigma}}{d\hat{t}}(ab\rightarrow cd)$ is the differential cross section for the subprocess\cite{64,65} such as $q\bar{q}\rightarrow Q\bar{Q}$, $qQ\rightarrow qQ$, $gQ\rightarrow gQ$, and $gg\rightarrow Q\bar{Q}$; $D_{H/c}(z_{c})$ is the Peterson heavy quark fragmentation function into charmed-meson\cite{66}, as well as $T_{A}(\mathbf{r})$ is the nuclear thickness function\cite{67,68} normalized to $\int d^{2}r T_{A}(\mathbf{r})=A$.

\begin{figure}
\includegraphics[width=8.5cm,height=7cm]{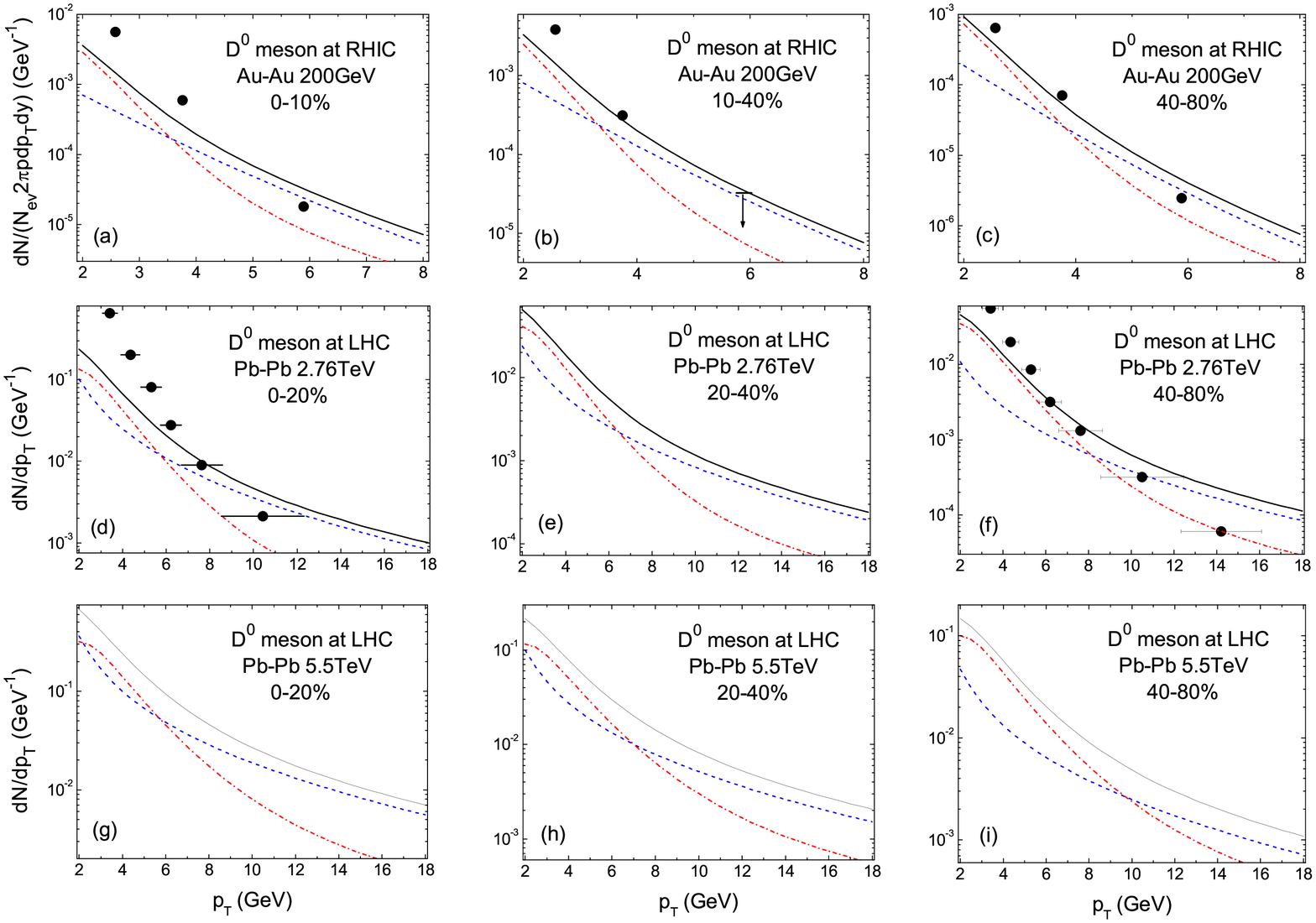}
\caption{\label{Fig1}  The centrality dependence of inclusive cross section of large transverse momentum $D^{0}$ meson production in the nucleus-nucleus collisions at RHIC and LHC energies. The $D^{0}$ meson data points are from the STAR Collaboration\cite{3} and ALICE Collaboration\cite{9}.}
\end{figure}

\begin{figure}
\includegraphics[width=8.5cm,height=7cm]{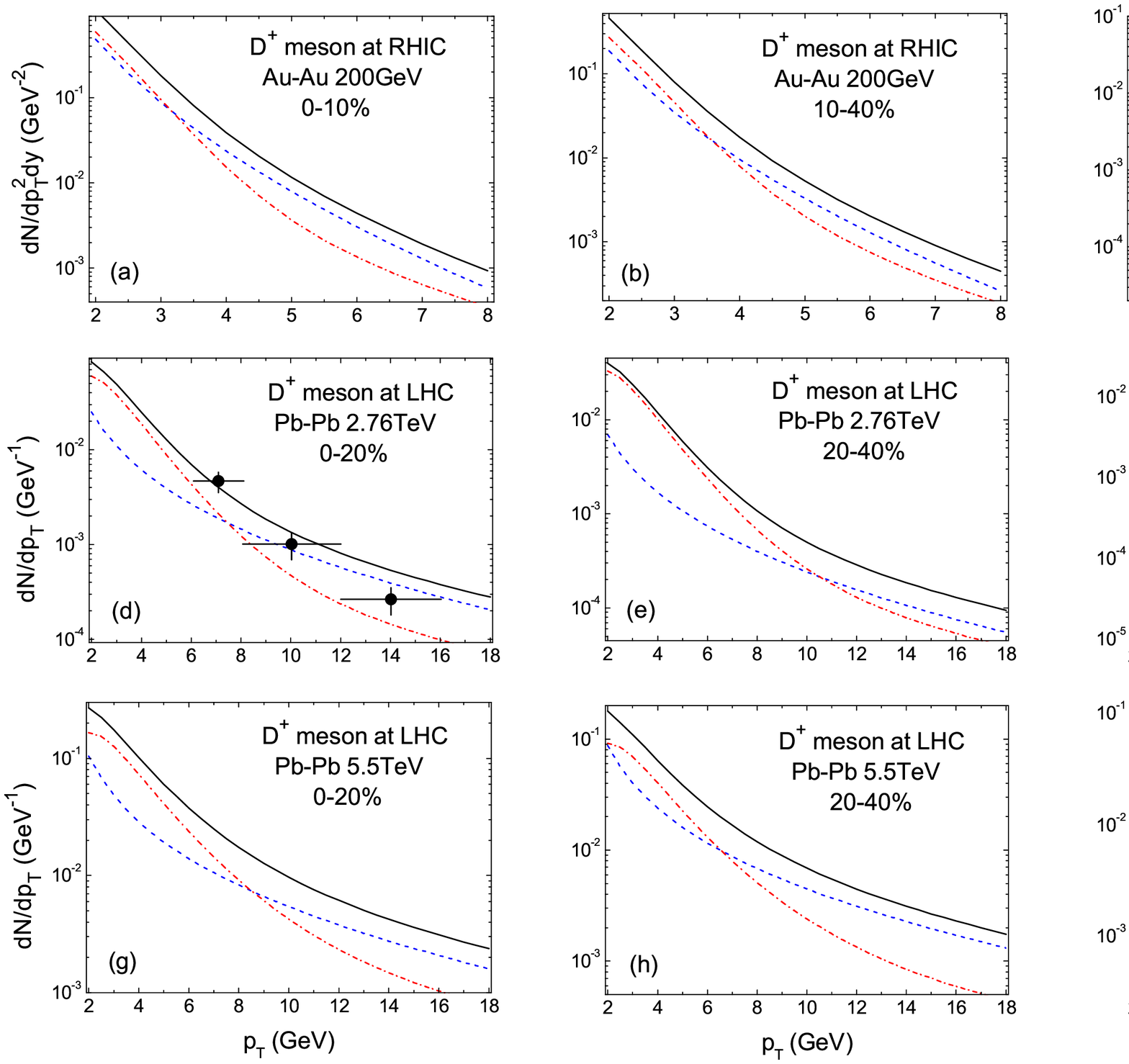}
\caption{\label{Fig2}  The centrality dependence of inclusive cross section of large transverse momentum $D^{*}$ meson production in the nucleus-nucleus collisions at RHIC and LHC energies. The $D^{*}$ meson data points are from ALICE Collaboration\cite{9}.}
\end{figure}

\begin{figure}
\includegraphics[width=8.5cm,height=7cm]{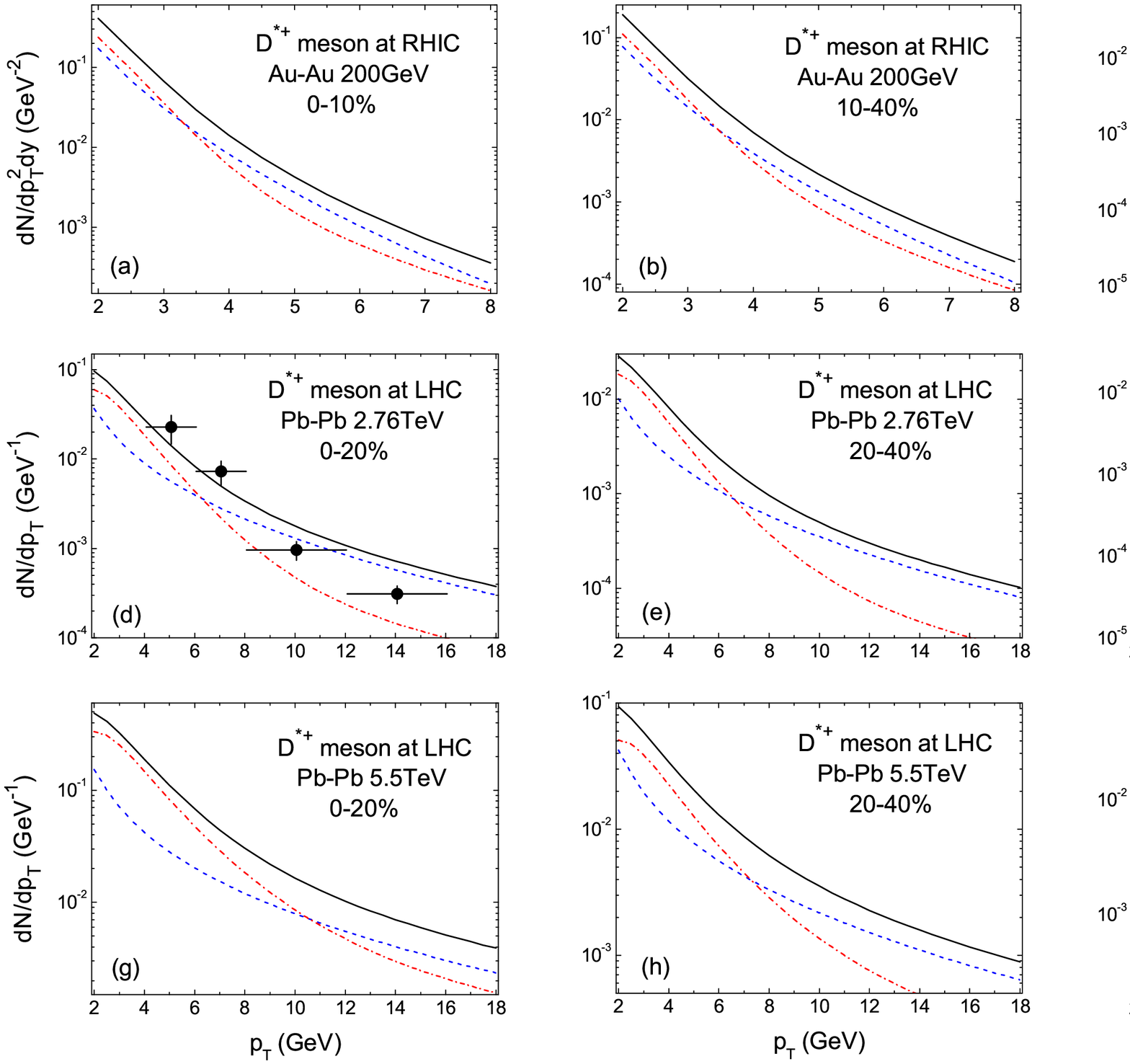}
\caption{\label{Fig3}  The centrality dependence of inclusive cross section of large transverse momentum $D^{*+}$ meson production in the nucleus-nucleus collisions at RHIC and LHC energies. The $D^{*+}$ meson data points are from ALICE Collaboration\cite{9}.}
\end{figure}

\begin{figure}
\includegraphics[width=8.5cm,height=7cm]{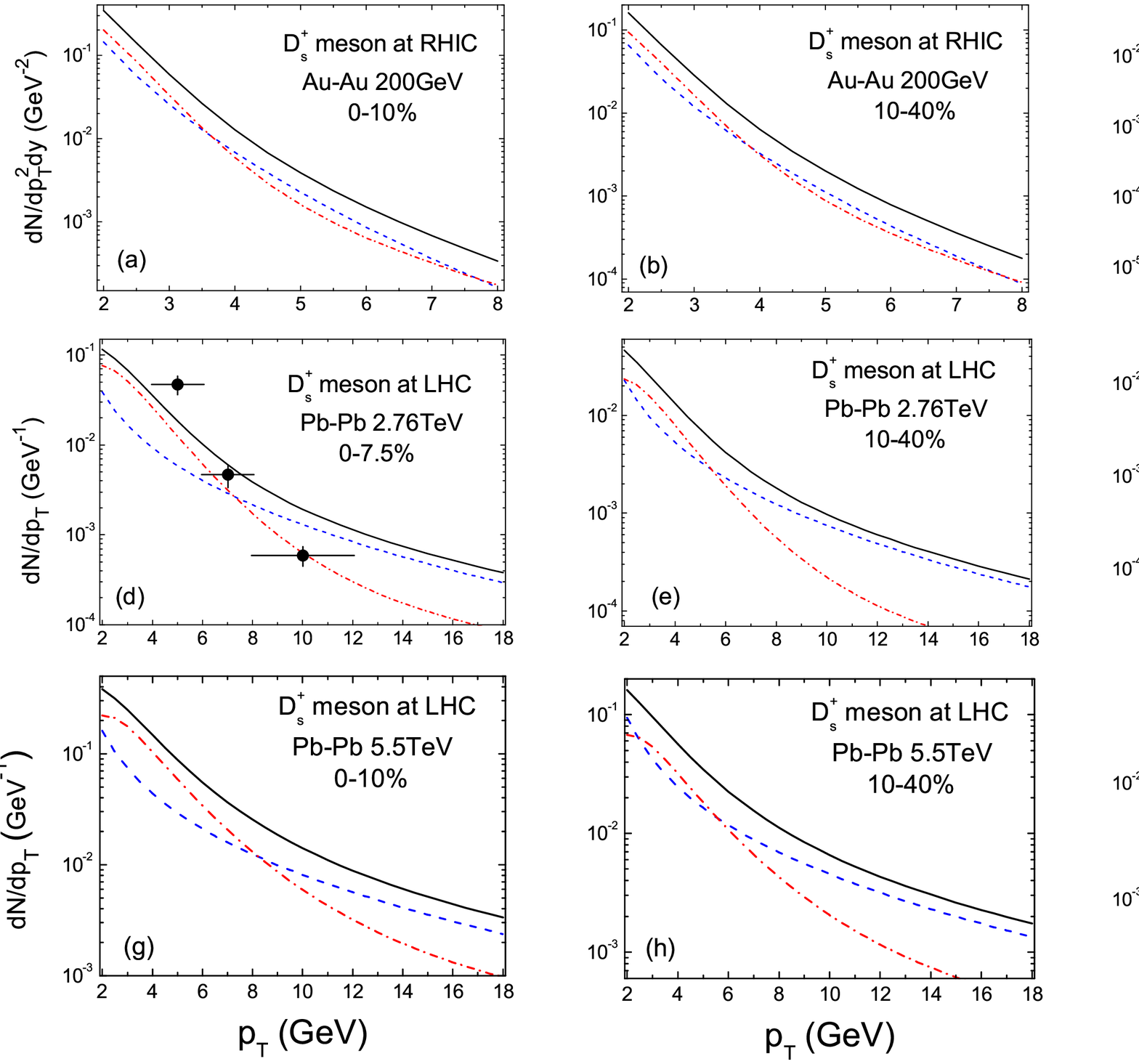}
\caption{\label{Fig4}  The centrality dependence of inclusive cross section of large transverse momentum $D_{s}^{+}$ meson production in the nucleus-nucleus collisions at RHIC and LHC energies.The $D_{s}^{+}$ meson data points are from ALICE Collaboration\cite{12}.}
\end{figure}

The parton distribution function $f_{a/A}(x,Q^{2},\mathbf{r})$ of the nucleon can be written as\cite{69}
\begin{eqnarray}\label{y2}
f_{a/A}(x,Q^{2},\mathbf{r})&=&R_{a/A}(x,Q^{2},\mathbf{r})\bigg[\frac{Z}{A}f_{a/p}(x,Q^{2})\nonumber\\[1mm]
&& +\left(1-\frac{Z}{A}\right)f_{a/n}(x,Q^{2})\bigg],
\end{eqnarray}
where $Z$ is the charge and $A$ the mass number of the nucleus. The parametrization\cite{70} for parton distribution functions will be used for nucleon parton distributions $f_{a/N}(x,Q^{2})$. The parton shadowing factor $R_{a/A}(x,Q^{2},\mathbf{r})$ describes the nuclear modification of parton distributions per nucleon inside the nucleus and can be given by the parametrization\cite{71,72,73,74}.

The large-$p_{T}$ charmed-meson produced by inelastic hard photoproduction processes can be divided into the inelastic direct hard photoproduction processes and inelastic resolved hard photoproduction processes in ultrarelativistic heavy ion collisions.

In the inelastic direct hard photoproduction processes, the charged parton of the incident nucleus can emit a photon, then the high energy photon interacts with parton of another incident nucleus by the gluon-photon fusion interaction. The differential cross section of inclusive large-$p_{T}$ charmed-meson produced by the inelastic direct hard photoproduction processes (inel.dir.) in the hadronic collisions can be expressed as
\begin{eqnarray}\label{y3}
\frac{dN^{inel.dir.}}{dp_{T}^{2}dy}&=&\!\!\!\int d^{2}bd^{2}rdx_{a}dx_{b}dz_{a}T_{A}(\mathbf{r})T_{B}(|\mathbf{r}-\mathbf{b}|)\nonumber\\[1mm]
&&\!\!\!\!\times f_{a/A}(x_{a},Q^{2},\mathbf{r})f_{\gamma/q}(z_{a})f_{b/B}(x_{b},Q^{2},|\mathbf{r}\!-\!\mathbf{b}|)\nonumber\\[1mm]
&&\!\!\!\!\times \frac{x_{a}x_{b}z_{a}}{x_{a}x_{b}z_{a}-\tau}\frac{d\hat{\sigma}}{d\hat{t}}(\gamma g\rightarrow Q\bar{Q})\frac{D_{H/c}(z_{c})}{z_{c}},
\end{eqnarray}
where $\frac{d\hat{\sigma}}{d\hat{t}}(\gamma g\rightarrow Q\bar{Q})$ is the differential cross section for the subprocess\cite{75} and the equivalent photon spectrum function of the charged parton is given by\cite{76,77}
\begin{eqnarray}\label{y4}
\!\!\!\!\!\!\!\!f_{\gamma/q}(x)\!&=&\!\frac{\alpha}{2\pi}e_{f}^{2}\bigg\{\frac{1+(1-x)^{2}}{x}\left[\ln\left(\frac{E}{m}\right)-\frac{1}{2}\right]\nonumber\\[1mm]
&&\!\!+\!\frac{x}{2}\left[\ln\left(\frac{2}{x}\!-\!2\right)\!+\!1\right]\!+\!\frac{(2\!-\!x)^{2}}{2x}\ln\!\!\left(\frac{2\!-\!2x}{2\!-\!x}\right)\!\!\bigg\},
\end{eqnarray}
here $x$ is the photon momentum fraction. The variables $e_{f}$, $E$, and $m$ are the charge, energy, and mass of the parton, respectively.

In the inelastic resolved hard photoproduction processes, the parton from the hadron-like photon emitted by the charged parton of incident nucleus can interact with the parton of another incident nucleus via the interactions of quark-antiquark annihilation, quark-gluon Compton scattering, and gluon-gluon fusion. The invariant cross section for inclusive large-$p_{T}$ charmed-meson produced by the inelastic resolved photoproduction processes (inel.res.) in the hadronic collisions can be written as
\begin{eqnarray}\label{y5}
&&\frac{dN^{inel.res.}}{dp_{T}^{2}dy}=\int d^{2}bd^{2}rdx_{a}dx_{b}dz_{a}dz'_{a}T_{A}(\mathbf{r})T_{B}(|\mathbf{r}-\mathbf{b}|)\nonumber\\[1mm]
&&\times f_{a/A}(x_{a},Q^{2},\mathbf{r})f_{\gamma/q}(z_{a})f_{\gamma}(z'_{a},Q^{2})f_{b/B}(x_{b},Q^{2},|\mathbf{r}-\mathbf{b}|)\nonumber\\[1mm]
&&\times \frac{x_{a}x_{b}z_{a}z'_{a}}{x_{a}x_{b}z_{a}z'_{a}-\tau}\frac{d\hat{\sigma}}{d\hat{t}}(\gamma g\rightarrow Q\bar{Q})\frac{D_{H/c}(z_{c})}{z_{c}},
\end{eqnarray}
where $z'_{a}$ is the momentum fraction of the parton from the hadron-like photon, and $f_{\gamma}(z'_{a},Q^{2})$ is the parton distribution function of the
hadron-like photon\cite{78}.



In ultrarelativistic heavy ion collisions, the equivalent photon
spectrum for the charged parton is $f_{\gamma/q}\propto\ln(E/m_{q})=\ln(\sqrt{s}/2m_{q})+\ln(x)$, where $m_{q}$ is the charged parton mass. Since the collision
energy $\sqrt{s}$ at RHIC and LHC are very large, the photon spectrum becomes important. Hence the photon spectrum for the charged parton becomes prominent at RHIC and LHC energies. Therefore the contribution of inclusive large-$p_{T}$ charmed-meson produced by photoproduction processes is evident at LHC. The numerical results of our calculation for large-$p_{T}$ charmed-meson produced by the hard photoproduction processes in relativistic heavy ion collisions are plotted in Figs.\,\ref{Fig1}-\ref{Fig4}. Compared with the the initial parton hard scattering processes (the dashed line) and the data of STAR Collaboration\cite{3} and ALICE Collaboration\cite{9,12}, we find that the large-$p_{T}$ charmed-meson produced by the hard photoproduction processes (the dashed-dotted line) cannot be negligible in the nucleus-nucleus collisions at RHIC and LHC energies.


\section{Conclusion}

We have investigated the inclusive production of large-$p_{T}$ charmed-meson from the direct and resolved hard photoproduction processes in Au-Au collisions at Relativistic Heavy Ion Collider (RHIC) and Pb-Pb collisions at Large Hadron Collider (LHC). At the early stage of relativistic heavy ion collisions, the charged parton of the incident nucleus can emit high photons, then the high energy photons interact with the partons of another incident nucleon by the photon-gluon fusion interactions. Furthermore, the partons of the hadron-like photons can interact with the partons of the nucleus by the quark-antiquark annihilation, quark-gluon Compton scattering, and gluon-gluon fusion scattering interactions. The numerical results indicate that the contribution of charmed-meson produced by the hard photoproduction processes cannot be negligible in the nucleus-nucleus collisions at RHIC and LHC energies.


\section{Acknowledgements}

This work is supported by the National Basic Research Program of China (973 Program) with Grant No.
2014CB845405 and the China Postdoctoral Science Foundation funded project with Grant No. 2017M610663.







\bibliographystyle{model1-num-names}
\bibliography{<your-bib-database>}

\end{document}